\documentclass[letterpaper, 10 pt, conference]{ieeeconf}  

\IEEEoverridecommandlockouts                              
\usepackage{cite}
\usepackage{amsmath,amssymb,amsfonts}
\usepackage{algorithmic}
\usepackage{graphicx}
\usepackage{textcomp}
\usepackage{xcolor}

\usepackage{algorithmic}
\usepackage{algorithm} 
\usepackage{bm}
\usepackage{color}
\usepackage{acronym}
\usepackage{rotating}
\usepackage{subfigure}
\usepackage{multirow}
\usepackage{colortbl}

\usepackage{textcomp}
\usepackage{graphicx}
\usepackage{makecell}
\usepackage{xcolor}
\usepackage{booktabs}
\usepackage{siunitx}
\usepackage{amsmath}
\usepackage{hyperref}
\usepackage{ulem}

\title{\LARGE \bf
Spike-Kal: A Spiking Neuron Network Assisted Kalman Filter
}

\author{Xun Xiao$^{1}$, Junbo Tie$^{1}$, Jinyue Zhao$^{2}$, Ziqi Wang$^{1}$, Yuan Li$^{1}$ \\Qiang Dou$^{3*}$ and Lei Wang$^{2*}$
\thanks{This work was supported in part by the National Natural Science Foundation of China under Grants 62372461, 62032001 and 62203457, in part by National Defense Key Laboratory Project of the State Administration of Science and Industry for National Defense under Grant WDZC20235250112 and in part by the Key Laboratory of Advanced Microprocessor Chips and Systems. (Corresponding author: Qiang Dou and Lei Wang)}
\thanks{$^{1}$Xun Xiao, Junbo Tie, Ziqi Wang, Yuan Li, Qiang Dou are with National University of Defence Technology, Changsha 410071, Hunan, P.R.China. E-mail:
        {\tt\small xiaoxun520@nudt.edu.cn}}%
\thanks{$^{2}$Jingyue Zhao, Lei Wang are with Defense Innovation Institute, AMS. E-mail:
        {\tt\small leiwang@nudt.edu.cn}}%
}

\begin{document}
\maketitle
\thispagestyle{empty}
\pagestyle{empty}

\begin{abstract}
Kalman filtering can provide an optimal estimation of the system state from noisy observation data. 
This algorithm's performance depends on the accuracy of system modeling and noise statistical characteristics, which are usually challenging to obtain in practical applications. 
The powerful nonlinear modeling capabilities of deep learning, combined with its ability to extract features from large amounts of data automatically, offer new opportunities for improving the Kalman filter. 
This paper proposes a novel method that leverages the Spiking Neural Network to optimize the Kalman filter. 
Our approach aims to reduce the reliance on prior knowledge of system and observation noises, allowing for adaptation to varying statistical characteristics of time-varying noise. 
Furthermore, we investigate the potential of SNNs in improving the computational efficiency of the Kalman filter. 
In our method, we design an integration strategy between the SNN and the Kalman filter. 
The SNN is trained to directly approximate the optimal gain matrix from observation data, thereby alleviating the computational burden of complex matrix operations inherent in traditional Kalman filtering while maintaining the accuracy and robustness of state estimation. 
Its average error has been reduced by 18\%-65\% compared with other methods.
\end{abstract}

\section{introduction}
Estimating the hidden state of a dynamical system from noisy observations in real time is a fundamental task in control and navigation. In the early 1960s, Kalman filter\cite{kalman1960new} was proposed to solve this problem. It can provide optimal estimates in the presence of observation noise. The core principle of Kalman filtering is using the current observation data combined with the previous state estimation to update the estimation of the current state. It is relatively simple to implement, as it does not require processing the entire history of observations but instead updates the current state. So that, this approach has been extensively applied across various fields of engineering, including navigation \cite{zhang2022new,hu2023robust,wang2023system}, control systems \cite{ahn2009online,ampountolas2023unscented}, signal processing \cite{roth2017ensemble,khodarahmi2023review}, and robot motion planning \cite{zhu2019motion,guo2019overhead,kang2020asie} et al.

However, the performance of Kalman filter is highly dependent on the accuracy of modeling.
Only by making accurate estimates of the system state and observed values can the optimal estimation be obtained.
On the other hand, the implementation of Kalman filters relies on matrix inversion and multiplication operations, especially in the calculation of gain matrices.
As the system dimension increases, the computational overhead of these operations increases exponentially.
The computational requirements of Kalman filter make it difficult to apply in scenarios involving high-dimensional systems or scenarios with limited computing resources.

In recent years, the emergence of neural networks has provided new insights for the improvement of Kalman filters.
Neural networks can approximate complex functions and distinguish complex patterns from data, providing a promising approach for accelerating and improving the computation of Kalman filters\cite{abiodun2018state}.
In particular, Spiking Neural Network (SNN), as a distinctive implementation of neural networks, has garnered widespread attention\cite{ghosh2009spiking}.
Inspired by biophysical processes in biological neurons, SNN encodes information in discrete spikes that are only computed upon receiving a spike. This computing mode achieves high energy efficiency and parallelism in information processing.
In addition, due to the continuity of membrane potential accumulation along the temporal dimension, each neuron can be regarded as a dynamic system.
The network made of such neurons has a powerful ability in memorize and detect spatiotemporal information.
These characteristics endow SNN with significant advantages in temporal data processing and energy-efficient computing, promising to offer new solutions for optimizing Kalman filters.

Some work has already used SNN to assist in Kalman filtering calculations. Ju{\'a}rez-Lora et al. \cite{juarez2022implementation} use SNN and R-STDP learning algorithms to calculate the gain matrix during the state update process of the Kalman filter. However, this method has a relatively slow convergence speed. Dethier et al. \cite{dethier2011spiking}mapped a Kalman filter based prosthetic decode algorithm developed to predict the arm's velocity onto the SNN using the Neural Engineering Framework\cite{stewart2012technical} and simulated it in Nengo\cite{bekolay2014nengo}. However, this method requires highly heterogeneous spiking neurons to achieve nonlinear multidimensional vector state transitions, which will incur substantial computational overhead.

In this paper, we explore integrating SNN into the Kalman filter framework, focusing on calculating the gain matrix, a key component in the filter update step. 
Our approach aims to reduce the reliance on prior knowledge of system and observation errors, allowing for adaptation to varying statistical characteristics of time-varying noise.
In addition, we mapped the designed filtering framework on our developed neuromorphic processor to verify its effectiveness.
The result show that its average error has been reduced by 18\%-65\% compared with other methods.

The main contributions of this paper are:
\begin{itemize}
\item[$\bullet$]
We have developed Spike-Kal, an interpretable, low-complexity SNN-aided filter. Spike-Kal builds upon the theoretical foundation of the Kalman filter, utilizing biologically spiking neuron and synapse models.

\item[$\bullet$]
We have optimized the model architecture and training method, significantly improving the convergence speed and filtering performance compared to existing SNN-assisted Kalman filters.

\item[$\bullet$]
We have thoroughly evaluated our methodology in various scenarios. The experimental scenarios include Linear motion, Lorentz system, and UAV trajectory measurement in real-world scenarios. Meanwhile, we deployed our method on a neuromorphic processor and verified its feasibility in real-world scenarios.
\end{itemize}

\section{background and related work}
\subsection{State Space Model and Kalman Filter}
The state space model is a mathematical framework for representing dynamic systems.
It represents the system's state at different times via state variables and describes the evolution and observation of the system's state using the state equation and observation equation, respectively. 
We consider a linear state space model with Gaussian noise, which can be expressed as follows:
\begin{equation}
\mathbf{x}_{t} = \mathbf{Ax}_{t-1} + \mathbf{w}_{t}, \mathbf{w}_t \sim \mathcal{N}(\mathbf{0}, \mathbf{Q})
\end{equation}
\begin{equation}
\mathbf{y}_{t} = \mathbf{Hx}_{t} + \mathbf{v}_{t}, \mathbf{v}_t \sim \mathcal{N}(\mathbf{0, R})
\end{equation}

Where $\mathbf{x}_t$ is the state vector of the system at time $t$, which evolves from the previous state $\mathbf{x}_{t-1}$, by a linear state evolution matrix $\mathbf{A}$ and by a Gaussian noise $\mathbf{w}_t$ with covariance $\mathbf{Q}$.
$\mathbf{y}_t$ is the state vector of observations at time $t$, which is generated from the current latent state vector by a linear observation $\mathbf{H}$ corrupted by a Gaussian noise $\mathbf{v}_t$ with covariance $\mathbf{R}$.

The Kalman filter can dynamically iterate to obtain an optimal solution from noisy system state evolves and imperfect observation. The core of a Kalman filter is to weigh the average of two sources of information: the predicted state obtained from system dynamics and the observed values obtained from external sensors. Both types of information are inherently affected by noise and inaccuracy, while information with smaller uncertainties will be assigned larger gain coefficients.
The state space can be modeled using the following equations:

In the prediction stage, the Kalman filter utilizes the dynamic model of the system and previous state estimates to predict the next state of the system while updating the covariance matrix.
\begin{equation}
\mathbf{\hat{x}}_{t} = \mathbf{Ax}_{t-1}
\label{1}
\end{equation}
\begin{equation}
\mathbf{\hat{P}}_{t} = \mathbf{AP}_{t-1}\mathbf{A^T} + \mathbf{Q}
\label{2}
\end{equation}

In the update phase, the Kalman filter combines new measured values with predicted states to refine state estimation. The essence of this stage lies in the calculation of the Kalman gain matrix, which is used to weight the predicted state optimally and observed values, thereby correcting measurement noise and system uncertainty.
\begin{equation}
\mathbf{K} = \frac{\mathbf{\hat{P}}_{t}\mathbf{H^T}}{\mathbf{H\hat{P}}_{t}\mathbf{H^T+R}}
\label{kcalculate}
\end{equation}
\begin{equation}
\mathbf{x}_{t} = \mathbf{\hat{x}}_t+\mathbf{K}(\mathbf{y}_t-\mathbf{H\hat{x}}_t)
\label{4}
\end{equation}
\begin{equation}
\mathbf{P}_{t} = \mathbf{(I-kH)\hat{P}}_t
\label{5}
\end{equation}

\subsection{Spiking Neural Network}
SNN represent a class of neural networks inspired by the spatiotemporal dynamics observed in biological neurons. Unlike traditional artificial neural networks, which operate based on continuous-valued activations, SNNs encode information in the form of discrete spikes, mirroring the asynchronous firing patterns observed in real neurons. A single neuron accumulates membrane potential upon receiving spike input and emits new spikes when the membrane potential reaches a threshold. Neurons form a complex network structure through synaptic connections. Neurons are the basic building blocks of snn. Many neuron models have been proposed in the literature, such as HH, Izhikevich, LIF, IF, etc. Among these models, the most commonly used is the LIF model. Due to its good biological rationality and low computational complexity, it can be described by the following formula:

\begin{equation}
    \frac{dV}{dt} =\frac {-(V - V_{rest})}{\tau_1}+\frac{I}{\tau_2}
\end{equation}

Where $V$ is the membrane potential and $I$ is the input received by the neuron. $V_{rest}$ is the resting potential, which is the membrane potential to which a neuron slowly leaks when it has not received any input pulses for an extended period. $\tau$ is the membrane's temporal charging constant. When a neuron receives a spike input, $I$ increases according to the strength of the synapse and continues to decay over time. The calculation of $I$ is updated as follows.

\begin{equation}
\frac{dI}{dt} = -\frac{I}{\tau_3}+ \sum W_i \cdot s_i
\end{equation}

Where $W$ is the synaptic weight and $s$ is the spike received by the neuron. 
\subsection{Related Works}
Kalman filter is the optimal estimator When the dynamic model and observation model of the system are linear and the noise distribution is Gaussian noise.
However, Kalman filtering requires a linear Gaussian model. 
In the case of nonlinear systems, Larson et al. proposed a method to handle nonlinear systems with Taylor series expansion, known as the Extended Kalman Filter (EKF)\cite{larson1967application}.
Furthermore, Julier et al. proposed Unscented Kalman Filter (UKF) using the unscented transform to address this problem\cite{julier1997new}.
For more complex models when the noise cannot be modeled as Gaussian, Particle filter approximates the state probability distribution by utilizing a set of randomly sampled "particles"\cite{gordon1993novel}.
Nevertheless, these model-based algorithms rely on accurate model parameters, and a mismatch in the model can lead to a degradation in performance.

In recent years, many researchers have combined neural networks with Kalman filtering due to their powerful data processing capabilities. 

Bai et al. \cite{bai2020neuron} designed an improved Kalman filtering framework that uses an ANN to model the system dynamics. This method omits the modeling process of traditional Kalman filtering, reducing the impact of system modeling and assumed parameters on accuracy.

Revach et al. \cite{revach2022kalmannet} introduced KalmanNet. KalmanNet incorporates a Recurrent Neural Network (RNN) module within the Kalman filtering process, using the RNN to compute the Kalman gain matrix. This approach retains the data efficiency and interpretability of the classic algorithm while implicitly learning complex dynamics from data.

Ju{\' a}rez-Lora et al. \cite{juarez2022implementation} improved the Kalman filtering framework by combining biologically plausible neuron models with spike-time-dependent plasticity learning algorithms, using SNNs to estimate the values of the Kalman gain matrix. The proposed filtering framework was validated through simulations on some representative nonlinear systems.

Dethier et al. \cite{dethier2011spiking} implemented a Kalman filter-based decoder for Brain-Machine Interface (BMI) applications to achieve cortically controlled motor prostheses. Given the stringent power dissipation constraints to avoid cortex damage, the authors used the Neural Engineering Framework to implement the existing Kalman filter-based decoder in an SNN. This decoder was deployed on an ultra-low-power neuromorphic chip to decode neural signals from these intracortical implants. Online closed-loop BMI experiments with two rhesus monkeys demonstrated that the performance of the Kalman filter implemented using a 2000-neuron SNN was comparable to that of the Kalman filter implemented using standard floating-point techniques.

The emergence of neural networks has provided a new method for improving the performance of Kalman filtering. However, existing methods based on ANN and partial SNN have complex models. The SNN with simple models has limited performance.

\section{method}
In this section, We will introduce our method in detail. We combine an SNN with Kalman filter to construct a hybrid, efficient computational architecture called Spike-Kal. We first introduce the overall workflow of our method, followed by a detailed description of each module and the training method.

\subsection{Overall Workflow}
Spike-Kal is an adaptation of the model-based Kalman filter, we use SNN to perform the most complex gain matrix calculation in the processing flow. Similar to model-based methods, it is divided into a prediction step and an update step. The schematic diagram of the algorithm is shown in Figure 1. 

\begin{figure}[h]
	\centering
	\includegraphics[width=0.95\linewidth]{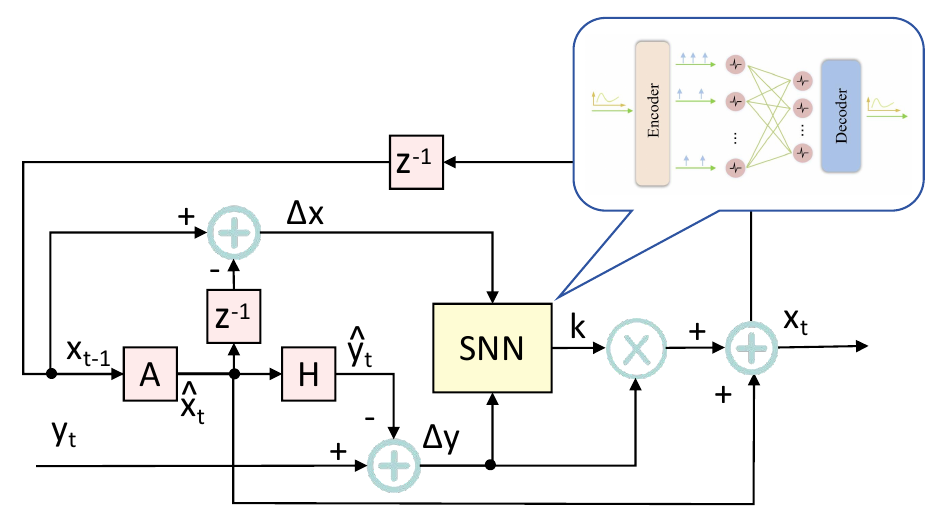}
	\caption{Spike-Kal block diagram.}
	\label{pic_DNF}
\end{figure}

In the prediction step, as shown in Equation 10, we use the posterior estimate of the system state from the previous time step to predict the prior estimate of the system state for the current time step. However, we do not explicitly estimate the second-order moments of the system state.
\begin{equation}
\mathbf{\hat{x}}_{t} = \mathbf{Ax}_{t-1}
\end{equation}

In the update step, we use the observation at the current time step to correct the prior state estimate based on the gain matrix, resulting in the posterior estimate of the system state for the current time step. Here, the gain matrix is not explicitly calculated through equations but is learned from the data using the SNN network. 

\begin{equation}
\mathbf{\Delta x} = \mathbf{x}_{t-1} - \mathbf{\hat{x}}_{t-1}
\label{deltax}
\end{equation}
\begin{equation}
\mathbf{\Delta y} = \mathbf{y}_{t} - \mathbf{H\hat{x}}_t
\label{deltay}
\end{equation}
\begin{equation}
\mathbf{K} = \mathbf{f(\Delta x,\Delta y)}
\end{equation}
\begin{equation}
\mathbf{x}_{t} = \mathbf{\hat{x}}_t+\mathbf{K(y}_t-\mathbf{H\hat{x}}_t)
\end{equation}

\subsection{Input Features}
The model-based Kalman filter computes the Kalman gain $\mathbf{K}$ by statistically evaluating the uncertainty of the system state and the observed state. 
As shown in Equation \ref{kcalculate}, the observation noise is explicitly provided at each iteration. 
At the same time, the uncertainty of the system state is iteratively calculated according to Equations \ref{2} and \ref{5}. 
In our method, we do not directly provide statistical characteristics of these two noises.
instead, we extract the information from the data using a neural network. 
We use two features as inputs for the network.
Feature 1 is the difference between the prior and posterior estimates of the system state from the previous iteration.
Feature 2 is the difference between the actual measurements of the model and the observed values.

When the filter has converged, $\mathbf{\hat{x}}_t$ can be approximately regarded as the true state of system, while $\mathbf{x}_t$ is the system's predicted value. Feature 1 thus carries the error between the system's predicted value and the true value, which represents the system noise.
Since we cannot obtain the result of the current time step, we use the result of the previous time step to approximate it.
Similarly, $\mathbf{y}_t$ is the true measurement value, and $\mathbf{\hat{x}}_t$ is the observation of the system state. Feature 2 thus contains both the system's uncertainty and the observation noise.

\subsection{Encoder and Decoder}
In our approach, we designed an encoding and decoding method for spiking data to connect traditional Kalman filter frameworks with SNN. 
Traditional Kalman filtering calculates in a continuous value, while SNN uses discrete spiking transmission and representation of data. 
The encoder module is responsible for converting state and measurement information into a form suitable for SNN processing. 
In contrast, the decoder module reconstructs the spiking signal output by SNN into continuous analog values for use by external systems.

There are currently many methods for converting analog inputs into spikes, the most commonly used being frequency coding, which uses the frequency of neuron spikes to represent the value of the input signal \cite{kiselev2016rate}. 
Generally, the longer the time step, the more accurate the restoration of input data. However, this method inevitably introduces errors. 
So, we directly use the simulated values of each time step as the current input for the neuron.

\subsection{Network Architecture}
\begin{figure}[h]
	\centering
	\includegraphics[width=0.95\linewidth]{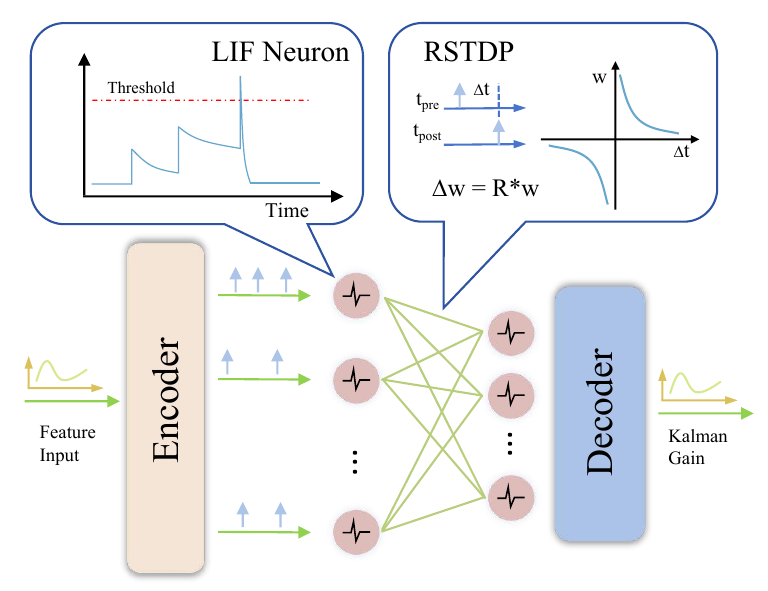}
	\caption{Spiking Neural Network and R-STDP.}
	\label{snn}
\end{figure}

Our proposed method features the SNN network as its core part, which replaces traditional Kalman filtering formula calculations. 
As shown in Fig.\ref{snn}, we employ a two-layer fully connected SNN network with R-STDP learning rules. 
The input of the network consists of $\mathbf{\Delta x}$ and $\mathbf{\Delta y}$, which are calculated according to Equations \ref{deltax} and \ref{deltay}, combined into a column vector.

\vspace{-10pt}
\vspace{-10pt}

The number of input neurons is equal to the input vector's length, with each component of the vector corresponding to an input neuron. The number of output neurons equals the number of elements in the gain matrix K, with each output neuron corresponding to a component in K. The input and output neurons are connected through synapses, forming a fully connected network. Synaptic weights are updating through R-STDP learning algorithm.

\subsection{R-STDP}
STDP is an unsupervised learning method that learns synaptic weight based on the timing relationship between pre- and post-synaptic spikes. This mechanism is considered a fundamental basis for learning and memory formation in the brain and is an important learning mechanism in biological neurons. The weight adjustment is calculated according to the following formula:
\begin{equation}
    \Delta w = 
    \begin{cases} 
    A^+ \exp\left(-\frac{\Delta t}{\tau^+}\right) & \text{if } \Delta t > 0 \\
    -A^- \exp\left(\frac{\Delta t}{\tau^-}\right) & \text{if } \Delta t < 0
    \end{cases}
\end{equation}

$\Delta t = t_{post} - t_{pre}$is the time difference between the spikes of the postsynaptic and presynaptic neurons. $A^+$ and $A^-$ are the learning rates, $\tau^+$ and $\tau^-$ are the constants for Long-Term Potentiation (LTP) and Long-Term Depression (LTD). Some modifications to the STDP algorithm were described in [8] to introduce a teaching signal into STDP. These modifications are based on dopamine's modulation of synaptic learning ability, as observed in biological systems. The following formula can describe the R-STDP learning rule:
\begin{equation}
    \Delta E = 
    \begin{cases} 
    A^+ \exp\left(-\frac{\Delta t}{\tau^+}\right) & \text{if } \Delta t > 0 \\
    -A^- \exp\left(\frac{\Delta t}{\tau^-}\right) & \text{if } \Delta t < 0
    \end{cases}
\end{equation}
\begin{equation}
    \Delta w = R \cdot \Delta E
\end{equation}

Where R is the reward signal. By introducing the reward signal, R-STDP can achieve goal-directed learning, improving the efficiency and effectiveness of learning. The network can adapt to more complex tasks in various environments.

\subsection{Training Method}
Currently, most neural network training is conducted offline using labeled datasets. We propose a teacher-model-guided approach for network training. Specifically, during the execution of the algorithm, we dynamically update the network weights according to the R-STDP algorithm. However, in the initial stage of filtering, using a gain matrix computed directly from an untrained SNN may result in filter divergence. Therefore, we first use the standard Kalman filter for computation, updating the system state with the standard Kalman filter while simultaneously guiding the SNN network training and updating the decoder neuron parameters. 
Due to the use of standard Kalman filtering for guidance, we can obtain relatively accurate K for calculation in the initial stage of filtering, reducing the probability of filter divergence while improving convergence speed.
When the training is complete, the standard Kalman filter computation is halted, and the SNN network carries out subsequent Kalman gain computations.

\section{experiment and results}
In this section, we have provided a detailed introduction to our experimental setup and results.  
We have fully validated our method on simulation and practical experiments.
\subsection{Experiment Setup}
In the simulation, system noise and observation noise that satisfy Gaussian distribution will be added to the simulation trajectory to simulate system modelling and sensor errors. The original trajectory without observation noise is used as the ground truth. We simulated two representative systems: Linear motion and the Lorenz system. In the practical experiment, we used a UAV trajectory in a real scene. The observation trajectory of the UAV is obtained with a detection algorithm\cite{xiao2023fe} and the ground truth is manually annotated. Furthermore, We deploy our algorithm on a neuromorphic processor.

We use a homogeneous neuromorphic processor \cite{yang2023back} with an extended neuromorphic acceleration instruction set for hardware deployment. This neuromorphic processor builds upon the RISC-V open-source instruction set by adding specialized instructions to support neuromorphic computing.
Compared to other heterogeneous processors, this processor not only enables neuromorphic computing but also supports the execution of standard numerical and control instructions through its basic instruction set. It is highly adapted for networks that combine neuromorphic computation and numerical computations.

We used the Mean Absolute Error (MAE) as the performance evaluation metric. MAE quantifies the average of absolute differences between corresponding points in the trajectories, thereby reflecting the degree of deviation between the estimated values and the actual values. A lower MAE indicates a higher degree of similarity between the two trajectories. The calculation of MAE is shown in Equation \ref{mae}.
\begin{equation}
    MAE = \frac{1}{n} \sum_{i=1}^{n} |y_i - \hat{y}_i|
    \label{mae}
\end{equation}
where $y_i$ is the true value and $\hat{y}_i$ is the estimated value.

We compared our method with KF, EKF and SNN Kalman\cite{juarez2022implementation}. We simulated the scenario with partial information in which the covariance matrices Q and R used in the KF and EKF were not optimal.

\subsection{Linear Motion}
The first simulated dynamic system was a liner motion, with the system variables including position and velocity in both the horizontal and vertical directions. 
The horizontal and vertical positions are represented by $X$ and $Y$ respectively, and the velocity is represented by $V_x$ and $V_y$.
The system states are updated according to Equation \ref{liner}($\mathbf{w}_t \sim \mathcal{N}(\mathbf{0}, \mathbf{Q})$), while the noisy observations are obtained using Equation \ref{linerH}($\mathbf{v}_t \sim \mathcal{N}(\mathbf{0}, \mathbf{Q})$).
The time step of the simulation experiment was 10 ms, and the entire simulation lasted 30 seconds.

\begin{equation}
\label{liner}
    \begin{pmatrix}
    X \\
    Y \\
    V_{x} \\
    V_{y}
    \end{pmatrix}
    =
    \begin{pmatrix}
    1 & 0 & dt & 0 \\
    0 & 1 & 0 & dt \\
    0 & 0 & 1 & 0 \\
    0 & 0 & 0 & 1
    \end{pmatrix}
    \cdot
    \begin{pmatrix}
    X \\
    Y \\
    V_{x} \\
    V_{y}
    \end{pmatrix}
    +
    \mathbf{w}_t
\end{equation}

\begin{equation}
\label{linerH}
    \begin{pmatrix}
    X_{ob} \\
    Y_{ob}
    \end{pmatrix}
    =
    \begin{pmatrix}
    1 & 0 & 0 & 0 \\
    0 & 1 & 0 & 0
    \end{pmatrix}
    \cdot 
    \begin{pmatrix}
    X \\
    Y \\
    V_{x} \\
    V_{y}
    \end{pmatrix}
    +\mathbf{v}_t
\end{equation}

\begin{figure}[h]
	\centering
	\includegraphics[width=0.95\linewidth]{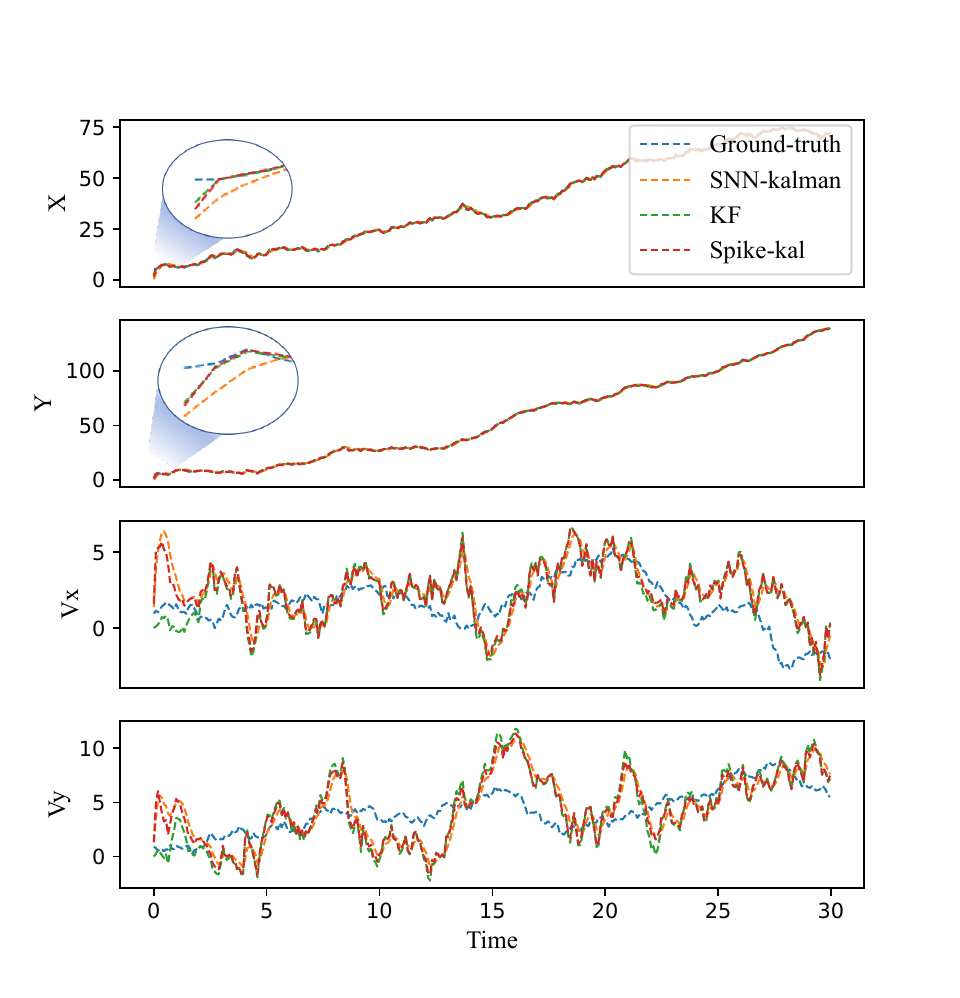}
	\caption{Comparison of reconstruction results of various methods on linear motion system with ground truth.}
	\label{liner}
\end{figure}
Fig.\ref{liner} shows the comparison of our method with the Kalman filter and SNN Kalman in scenarios of linear motion. 
Due to the position information can be directly observed, the position error is relatively smaller, while the reconstruct speed error is larger.
Comparing the convergence speed, our method is comparable to Kalman filter and superior to other SNN based methods.
\begin{table}[h]  \caption{Linear motion.}
    \centering
	\label{table_result1}
	\begin{tabular}{cccccc}
		\toprule   
		\multirow{2}{*}{Method}      & \multirow{2}{*}{Neurons}     &  \multicolumn{4}{c}{MSE} \\ 
        \cline{3-6}
		    && X     & Y& $V_x$& $V_y$\\  
		\midrule   
		KF   &-& 0.19 & 0.21 & \textbf{1.46} & 2.05\\  
		SNN kalman & 28 & 0.50 & 0.56 & 1.49 & \textbf{2.00} \\
		\midrule
		Spike-Kal(Ours) &14& \textbf{0.16} & \textbf{0.17} & 1.47 & 2.01\\
		\bottomrule 
	\end{tabular}
\end{table}

Table \ref{table_result1} shows the performance of our method with others. 
The results indicate that our method performs similarly to other methods in speed estimation, but reduces errors by 15\% -60\% in position estimation.

\subsection{Lorenz System}
The Lorenz attractor is a three-dimensional chaotic solution to the Lorenz system of ordinary differential equations in continuous time.
It is a typical example of a nonlinear system characterized by the following nonlinear dynamics, where $\mathbf{w}_t \sim \mathcal{N}(\mathbf{0}, \mathbf{Q})$.
The experiment simulated continuously for 30 seconds with a time step of 10ms.
\begin{equation}
    \label{}
    \begin{pmatrix}
    X_{1} \\
    X_{2} \\
    X_{3}
    \end{pmatrix}
    =
    \begin{pmatrix}
    -10 & 10 & 0  \\
    28 & -1 & -X_1  \\
    0 & X_1 & -8/3
    \end{pmatrix}
    \cdot
    \begin{pmatrix}
    X_1 \\
    X_2 \\
    X_{3}
    \end{pmatrix}
    +
    \mathbf{w}_t
\end{equation}

In this simulation experiment, we set the observation matrix to [1,0,0], which means that only the state $X_1$ Can be directly observed, while $X_2$ and $X_3$ need to be recovered by estimation. For nonlinear systems, we use Taylor series expansion to linearly approximate the model, with Extended Kalman filter as the comparison method
\begin{figure}[h]
	\centering
	\includegraphics[width=0.95\linewidth]{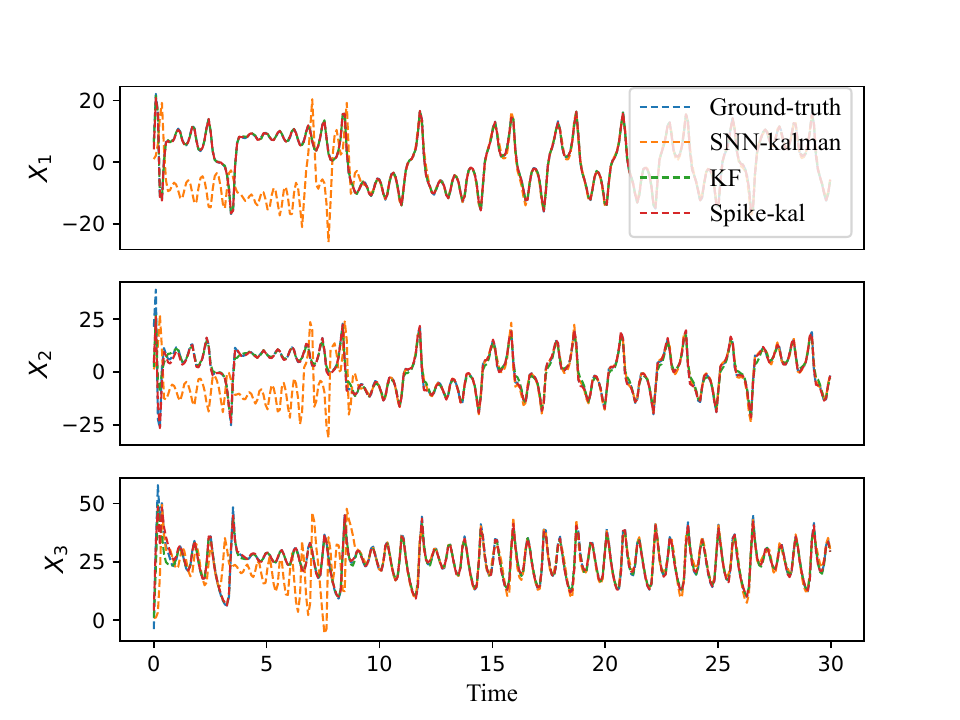}
	\caption{Reconstruction results of various methods on Lorenz system with ground truth.}
	\label{lorenzeresult}
\end{figure}

As shown in Fig.\ref{lorenzeresult}, both our method and the Extended Kalman filter can converge rapidly to the true values. However, the SNN Kalman only begins to converge after running for a period of time.
\begin{table}[h]  \caption{The MSE result of Lorenz system.}
    \centering
	\label{table_result}
	\begin{tabular}{ccccc}
		\toprule
		\multirow{2}{*}{Method} & \multirow{2}{*}{Neurons} &  \multicolumn{3}{c}{MSE}        \\
        \cline{3-5}
		   & & $X_1$ & $X_2$ & $X_3$ \\
		\midrule
            EKF & - &0.20 & 1.11 & 0.73  \\
            
		SNN kalman & 14 & 5.01 & 5.79 & 3.94 \\
		\midrule
		Spike-Kal(Ours) & 7 &  \textbf{0.16} & \textbf{0.48} & \textbf{0.60} \\
		\bottomrule
	\end{tabular}
\end{table}
\vspace{-10pt} 

Table \ref{table_result} shows more detailed information. The system variable $X_1$ has a lower error because it can be directly observed. Additionally, compared to the Extended Kalman filter, our method reduces the filtering error by 18\% to 65\%. Due to the lack of convergence of the initial trajectory, SNN Kalman's error is more larger than ours.

\subsection{Real World Experiment: UAV Tracking}
In our final experiment, we validated the effectiveness of our method in a practical scenario. We used Spike-Kal to track the trajectory of a UAV. We assume that the UAV is moving at a constant speed for a short period of time, so the state update of the filter is also modeled as Equation \ref{liner}.In this experiment, we could not obtain accurate information on the UAV's speed, so we only calculated the error between the observed position and its actual position. The units of horizontal and vertical in the figure are pixels and simulation time step is 33ms and lasts for 100s
\begin{figure}[h]
	\centering
	\includegraphics[width=0.95\linewidth]{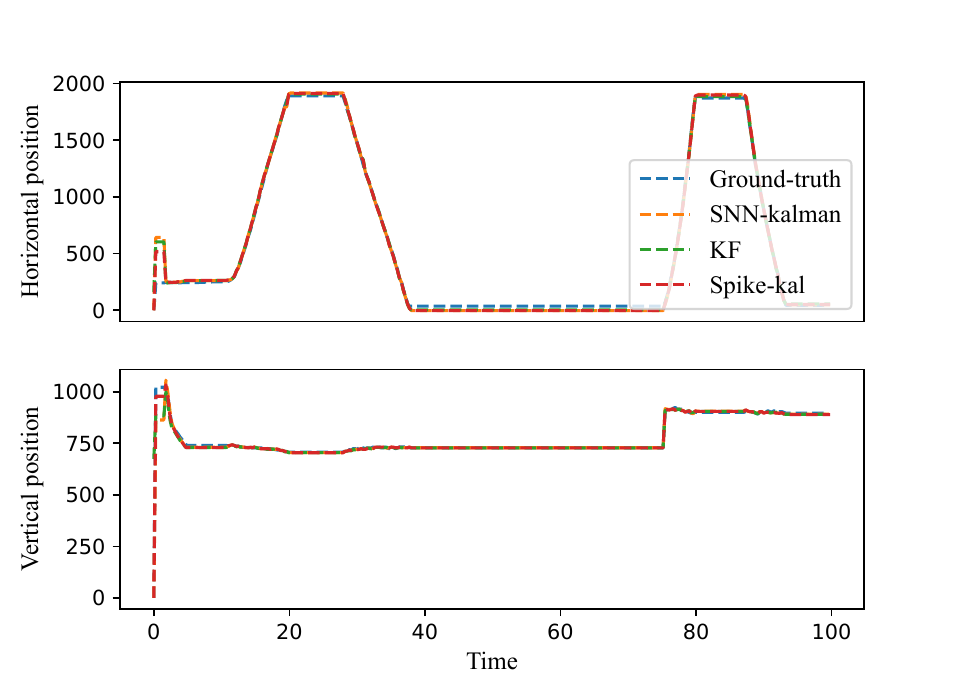}
	\caption{The results of UAV trajectory measurement.}
	\label{resultfig3}
\end{figure}
\vspace{-10pt} 
\begin{table}[h]  \caption{UAV tracking}
    \centering
	\label{table_result3}
	\begin{tabular}{cccc}
		\toprule
		\multirow{2}{*}{Method}      & \multirow{2}{*}{Neurons}     &  \multicolumn{2}{c}{MSE}        \\ 
        \cline{3-4} && Horizontal     & Vertical \\  
		\midrule
            KF & - & 8.67 & 4.64 \\
		SNN kalman & 28 & 10.15 & 5.00  \\
		\midrule
		Spike-Kal(Ours) &14& \textbf{8.10} & \textbf{3.24} \\
		\bottomrule
	\end{tabular}
\end{table}
\vspace{-10pt} 

As shown in Fig.\ref{resultfig3}, the trajectory obtained by our method is very close to the actual trajectory, and the standard Kalman filter and SNN Kalman filter are also roughly the same. The table\ref{table_result3} provides a more accurate evaluation by MSE. In the horizontal position, Our method has a 7\% and 20\% error reduction compared to KF and SNN kalman, respectively. In the vertical position, our method also achieves the best performance, reducing errors by 30\% to 35\%.
\section{conclusion}
In this paper, we have developed a novel Kalman filter by introducing an SNN. In the proposed method, SNN inspired us to optimize the intermediate computational processes in the Kalman filter using an R-STDP learning approach. This method primarily addresses the challenges of modeling and parameter estimation in traditional filters. Compared with existing methods, we significantly improved the convergence speed and filtering effect by introducing Kalman filtering to provide the Reward signal during the training phase Experiments were conducted on both simulated and real-world scenarios. 
The results show that our method can effectively filter data, with a performance improvement of 18\%-65\% compared to Kalman filters and other SNN based filters.
In addition, we implement our method on a neuromorphic process to verify our method. for future work, we will involve implementing more computations of the Kalman filter using SNN.

\addtolength{\textheight}{-11cm}   



\bibliographystyle{splncs04}·
\bibliography{software}
\end{document}